\documentclass{ws-p8-50x6-00}

\usepackage{wrapfig}
\usepackage{epsfig}

\def\be{\begin{equation}}
\def\ee{\end{equation}}
\def\bea{\begin{eqnarray}}
\def\eea{\end{eqnarray}}

\def\k0k0{\mbox{K}_s^0\mbox{K}_s^0}
\def\gg{\gamma \gamma}
\def\wgg{W_{\gamma \gamma}}
\def\ee{\mathrm{e^+ e^-}}
\def\mm{\mathrm{\mu^+ \mu^-}}
\def\gev{\mathrm{GeV^2}}
\def\jpsi{J/\psi}
\def\rar{\rightarrow}

\def\fg{F_{2}^{\gamma}}
\def\f2qed{F_{2}^{\gamma, QED}}
\def\f2qcd{F_{2}^{\gamma, QCD}}
\def\lgq2{lnQ^{2}}
\def\NPA{{\em Nucl. Phys.} A}
\def\PRC{{\em Phys. Rep.} C}
\def\EPJ{{\em Eur. Phys. J}}

\begin{document}

\title{Two Photon Physics at LEP}

\author{Maneesh Wadhwa}

\address{University of Basel, Klingelbergstrasse 82,\\ 
         CH-4056 Basel, Switzerland \\
         E-mail:Maneesh.Wadhwa@cern.ch}

\maketitle

\abstracts{LEP offers an excellent opportunity to measure two photon processes 
over a large kinematical range and thus study the complex nature of the photon. This
article reviews the experimental status of ``Two Photon Physics'' at LEP. The recent
results on resonances, multi-hadron production and photon structure functions are 
discussed.}

\section{Introduction}
\begin{wrapfigure}{r}{180pt}
\epsfig{file=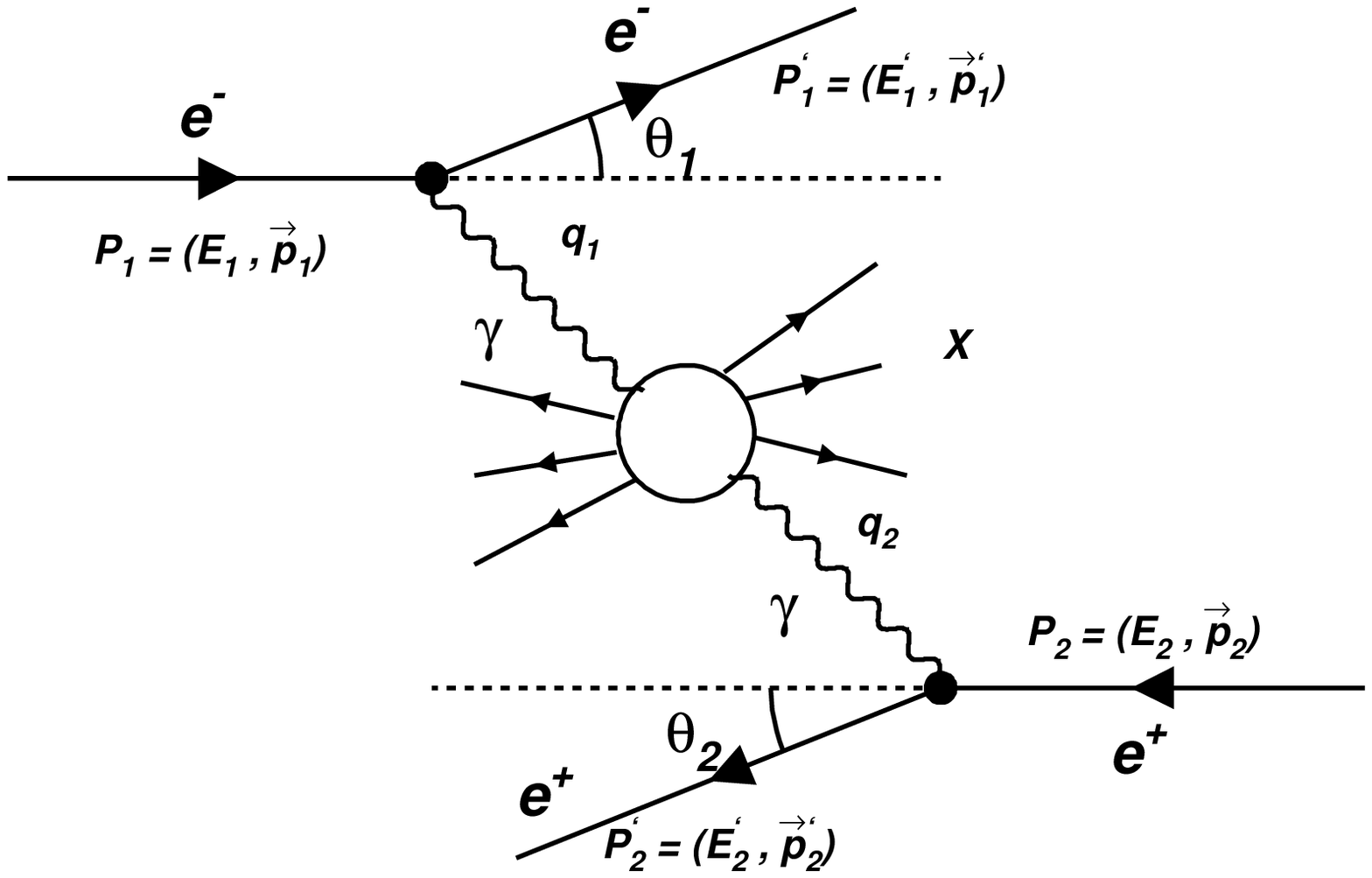,height=2.1in,width=2.5in}
\caption{\it $\gg$ collision in $\ee$ scattering}
\label{fig:feynman}
\end{wrapfigure}

Over the past decade two photon physics has proven to be a very productive source of
information about QED, QCD and hadron spectroscopy. The Feynman diagram responsible for 
a two photon collision process at LEP is shown in Figure~\ref{fig:feynman}, where the 
high energy incident electrons and positrons split off virtual photons and the 
scattered electrons take most of the beam energy.


\noindent These two photons then can interact to form a
state $X$ with  mass $\wgg$. The four-momentum transfer $q_{i}$ to the photons depends on the 
angle and energy of the scattered electrons\footnote{Electron stands for electron and 
positron throughout this article}. When neither of the scattered electrons is detected
(untagged events), the virtual photons are referred to as nearly real i.e. 
$q^{2}_{1} \approx q^{2}_{2} \approx 0$. This class of
events allows several tests of QCD by studying  hadronic resonances, the inclusive
hadron cross section and jet production rates. If there is detection of one of the 
scattered electrons $Q^2= -q^{2}_{1}$ (single tagged events), it is 
possible to probe the other photon $q^{2}_{2} \approx 0$ regarded as a ``target'' and 
study its structure. Finally, if both the scattered electrons are detected  
$Q^2_{i}= -q^{2}_{i}, (i=1,2)$ (double tagged events), the structure of the reaction of 
highly virtual photons is probed. In the following sections, a review is given of the 
$\gg$ results obtained at LEP, with special attention to recent results.

\section{Resonance production}
Two photon formation of C-even meson resonances provides valuable information on the internal
structure of mesons. In particular it is interesting to look for resonances whose $\gg$ 
couplings are much smaller than quark-model predictions; e.g. glueball or hybrid quark-gluon states.
One can also produce resonances in two-photon events in which one photon is far off mass shell.
The interest in this case is twofold. First, the meson transition form factor can be measured
and secondly spin-1 states can be produced.
\begin{table}[htp] 
\begin{center}
\caption{\it List of resonances studied at LEP\label{tab:table1}}
\begin{tabular}{|c|l|l|l|} \hline
Resonance              & Final state              & $\rm{J^{PC}}$&$\Gamma_{\gamma \gamma}$(keV)\\ \hline
$\eta^{'}~(~958) $     & $\pi^+ \pi^- \gamma    $ &   $0^{-+}$   &$4.17\pm 0.10 \pm 0.27$      \cite{etaprime} \\
$\rm{a_{2}}~(1320)$    & $\pi^+ \pi^- \pi^{0}   $ &   $2^{++}$   &$0.98\pm 0.05 \pm 0.09$      \cite{atwo}     \\
$\rm{a^{'}_{2}}~(1750)$& $\pi^+ \pi^- \pi^{0}   $ &   $2^{++}$   &$0.29\pm 0.04 \pm 0.02$/(BR) \cite{atwo}     \\
$\rm{f^{'}_{2}}~(1525)$&$\rm{K^{0}_{s} K^{0}_{s}}$&   $2^{++}$   &$0.09\pm 0.02 \pm 0.02$/(BR) \cite{ftwo}     \\
$\rm{\eta~(1440)}$     & $\rm{K^{0}_{s} K \pi}  $ &   $0^{-+}$   &$0.17\pm 0.05 $/(BR)         \cite{ftwot}    \\ 
$\rm{\eta_{c}~(2980)}$ & 12 Channels              &   $0^{-+}$   &$8.0~\pm 2.3~ \pm 2.4~$      \cite{etac1}    \\
$\rm{\eta_{c}~(2980)}$ &  9 Channels              &   $0^{-+}$   &$6.9~\pm 1.9~ \pm 2.0~$      \cite{etac2}    \\
$\rm{\chi_{c2}(3555)}$ & $\jpsi~\gamma          $ &   $2^{++}$   &$0.97\pm 0.40 \pm 0.36$      \cite{chic}     \\
\hline
\end{tabular}
\end{center}
\end{table}

\noindent At LEP, many exclusive channels are studied as shown in Table~\ref{tab:table1}. 
Two recent results are discussed in the following sections.

\subsection{Charmonium Production}

\begin{figure}[htp]
\begin{tabular}{cc}
 \includegraphics[width=13pc,height=11.3pc]{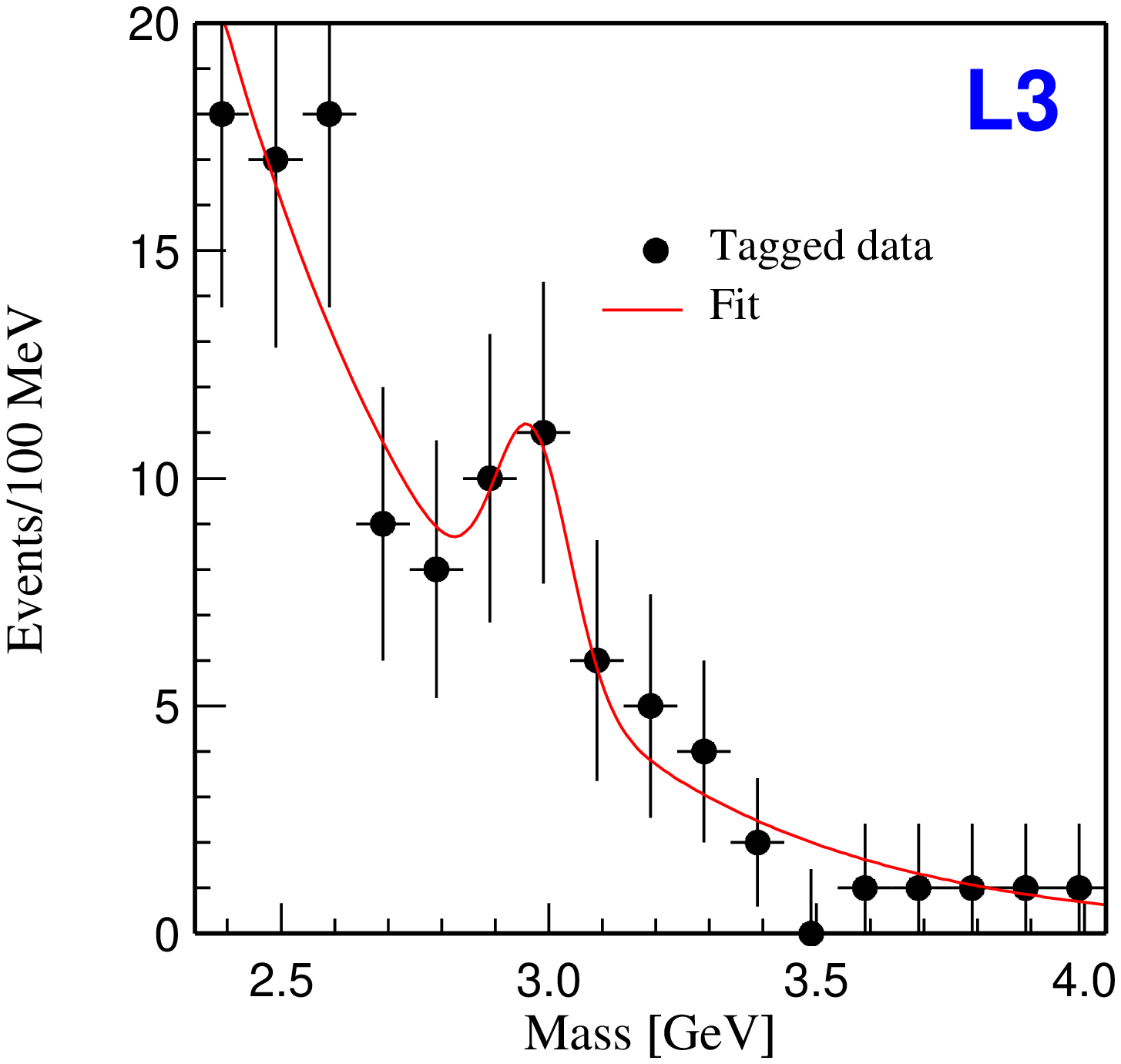}&
 \includegraphics[width=13pc,height=11.3pc]{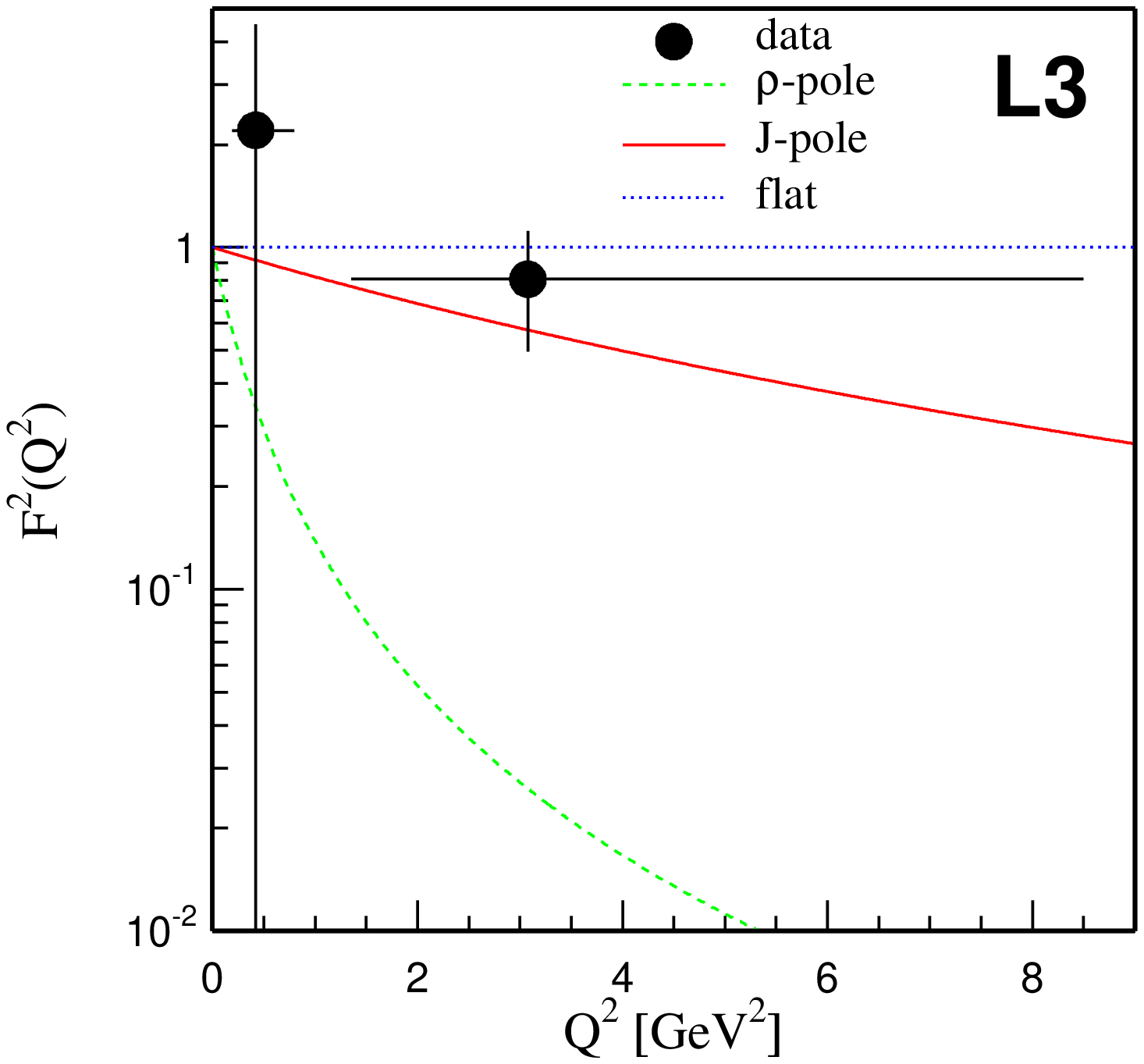} \\
(a) & (b)  \\
\end{tabular}
\caption{\it (a) The $\eta_{c}$ invariant mass spectrum , (b) the $\eta_{c}$ form factor, 
fitted with a VDM pole form, with pole mass equal to $\mathrm{M_{J/\psi}}$.}
\label{fig:etafig}
\end{figure}

Measurements of the charmonium system in the two photon collisions are mainly motivated by the
large quark mass, where the predictions are reliable, which provides a test of perturbative QCD.
Using LEP I and LEP II data, with a total luminosity of 193 $\rm{pb^{-1}}$ , the 
charmonium resonance $\eta_{c}$ is observed\cite{etac2} and reconstructed in 
nine different decay modes. The two photon partial width of the $\eta_{c}$ is 
extracted to be $\Gamma_{\gg} = 6.9 \pm 1.9 \pm 2.0 $~keV. Figure~\ref{fig:etafig} (a) 
shows the invariant mass distribution 
of selected events with one of the scattered electron tagged in the forward calorimeter. 
The spectrum is fitted with a Gaussian for the signal and a exponential for the background. 
These events allow to measure the $\eta_{c}$ transition form factor in different
$Q^{2}$ bins, (0.2~GeV$^{2}$ $<Q^{2} < 9$~GeV$^{2}$).  Figure~\ref{fig:etafig}(b) 
shows the $\eta_{c}$ form-factor measurement by L3, which favors the form-factor 
with a $\rm{J/\psi}$ mass pole in the VDM model and are in agreement with theoretical 
calculations\cite{pole}.

\subsection{$\k0k0$ Resonances and GlueBall Search }

The resonance formation process $\gamma\gamma \rar R \rar \k0k0\rightarrow\pi^+\pi^-\pi^+\pi^-$ 
has been
\begin{wrapfigure}[14]{r}{180pt}
\epsfig{file=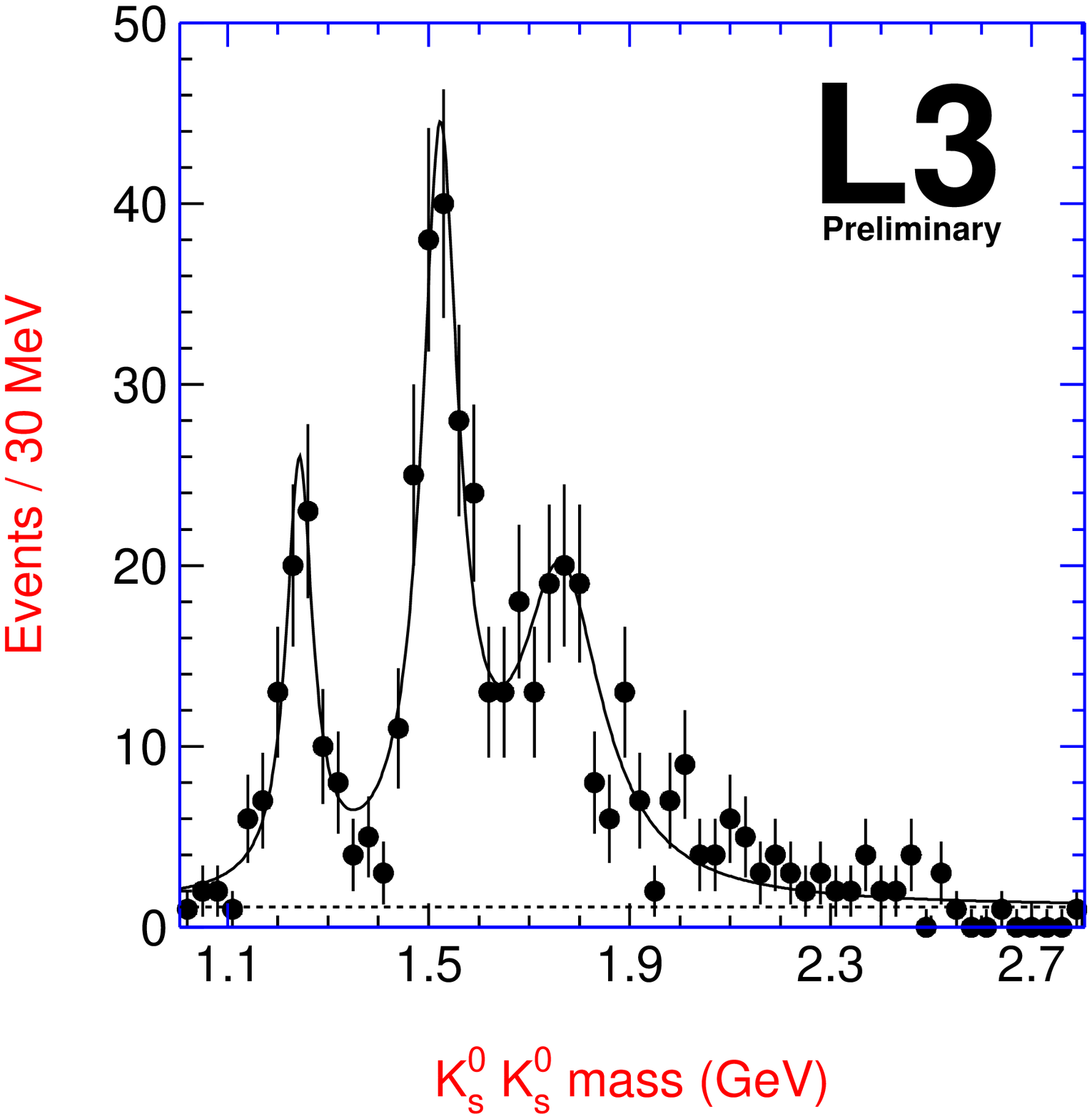,height=1.8in,width=2.5in}
\caption{\it The $\k0k0$ mass spectrum}
\label{fig:ksfig}
\end{wrapfigure}
\noindent studied\cite{ftwo} with the L3 detector. The $\k0k0$ mass spectrum  
Figure.~\ref{fig:ksfig}, shows clear evidence for the formation of the $f_2^{'}$(1525) tensor 
meson. Around 1300~MeV, $f_{2}(1270)-a_{2}(1320)$ destructive interference is observed 
consistent with theoretical predictions\cite{Lipkin}. In addition, there is an enhancement 
of $\approx$6 standard deviations around 1750 MeV which is possibly due to the formation of 
a radially excited state of the $f_2^{'}$, according to theoretical predictions\cite{Munz}. 
The measured two photon partial width of the ${f_{2}}^{'}(1525)$ is shown in Table~\ref{tab:table1}. 
A study of the angular distribution of the ${f_{2}}^{'}$  in the two-photon centre-of-mass 
system favours helicity-2 formation over helicity-0, consistent with theoretical predictions\cite{Kopp}. 

\noindent A search for the glueball candidate 
$\xi$ (2230) has been performed at LEP in the $\k0k0$ decay channel. The search is motivated due 
to the previous observation of $\xi$ (2230) by the Mark III Collaboration\cite{Mark} which has 
been confirmed by BES Collaboration\cite{BES}. At LEP, non observation of signal gives an upper 
limit for $\Gamma_{\gg}(\xi (2230)) \times \rm{Br( \xi (2230) \rightarrow \k0k0)} < 1.5$~eV at 
95\% CL under the hypothesis it is a pure spin 2, helicity two state. This low value is most 
likely inconsistent with a $\mathrm{q \bar{q}}$ assignment to the $\xi (2230)$.

\section{ The Two Photon Total Cross-section}

\begin{figure}[htp]
\begin{center}
\epsfig{file=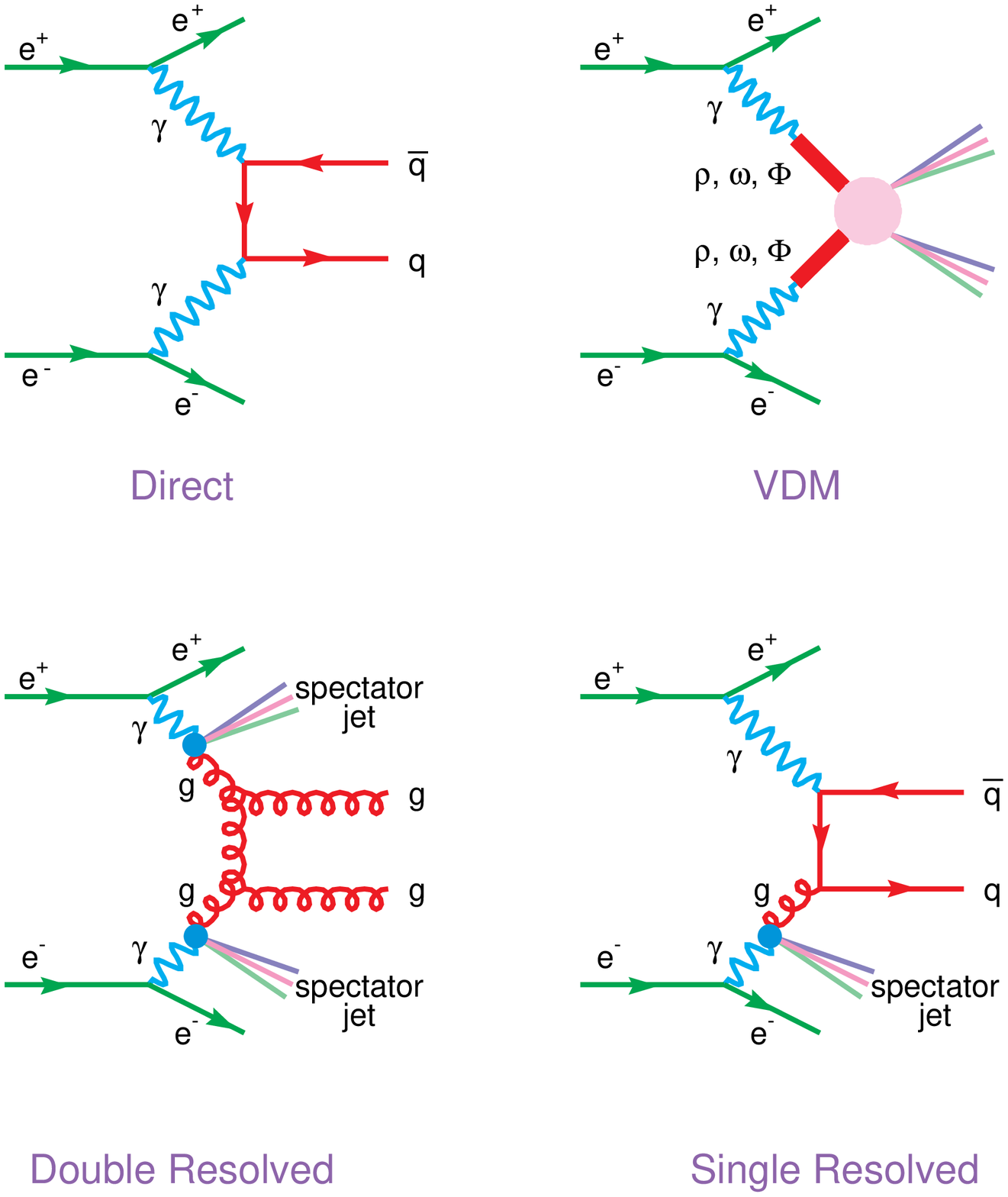,height=0.50\textwidth,width=0.55\textwidth} 
\caption{\it Some diagram contributing to hadron production in $\gg$ collisions at LEP.}
\label{fig:feyhad}
\end{center}
\end{figure}

At LEP II energies, the two photon process $e^+e^-\rar e^+e^- \gamma^* \gamma^* \rar e^+e^- hadrons$ 
is a copious source of hadron production. In this reaction the photons either interact as a 
point-like particle or undergo quantum fluctuation (resolved photon) into a resonant(VMD) or 
non-resonant virtual states opening up all the possibilities of hadronic interactions as shown 
in Figure~\ref{fig:feyhad}. These interactions can be described in terms of Regge poles\cite{regge,pomeron}, 
(Pomeron or Reggeon exchange). 

\begin{wrapfigure}{r}{180pt}
\epsfig{file=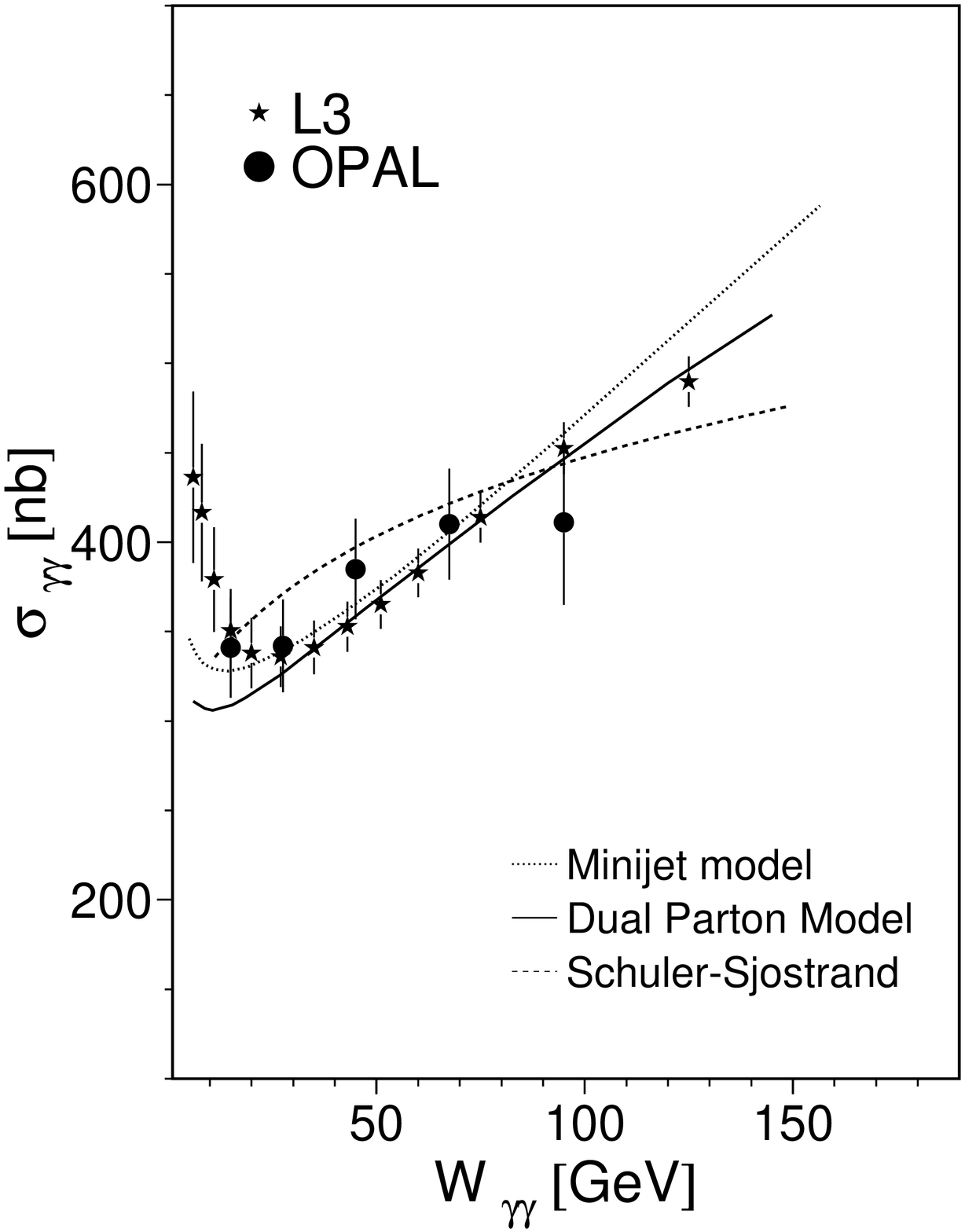,width=2.4in,height=2.6in} 
\caption{\it The measured cross-section $\sigma(\gg \rar hadrons)$ as a function of $\wgg$.}
\label{fig:hadron}
\end{wrapfigure}

\noindent A measurement of the total hadronic cross section as a 
function of $\sqrt {s} $, improves our understanding of the hadronic nature of the photon. At LEP, using
the high energy runs above the Z peak, L3 and OPAL have measured the cross section\cite{l3had,opalhad}
$\sigma(\gg \rar hadrons)$ in the range $5 \leq \wgg \leq 145$~GeV as shown in the 
Figure~\ref{fig:hadron}. 

\noindent The cross-section measurement of the two experiments show a clear rise at
high energies, described by a "Soft Pomeron" and the data of the L3 experiment show a fast decrease at 
low energies due to "Reggeon exchange".  The rise of $\sigma_{\gg}$ is faster than the one observed in 
hadron-hadron or $\gamma$p collisions; a simple factorization ansatz\cite{gribov} 
$\rm{\sigma_{\gg} = \sigma^{2}_{\gamma p}/\sigma_{pp}}$ is excluded as can be seen in 
Figure~\ref{fig:hadron} from the predictions of Schuler and Sjostrand\cite{modelsas}. The data are rather 
well described by the dual parton model of Engel and Ranft\cite{modeldual} or by analytical calculations 
which take into account the importance of QCD effects at high transverse momentum. In Figure~\ref{fig:hadron}, 
the minijet model of Godbole and Pancheri\cite{modelhad} is also represented. One has to notice
that all models has some dependence which can change the cross section predictions by 10-30\%.   
The Monte Carlo models PYTHIA and PHOJET which are used to correct the data, differ by 
$\approx$20\% in the absolute normalization. In future, improvements in the theoretical 
predictions especially the description of diffractive processes are desirable.


\section{Single Particle and Jet Production}

\begin{figure}[htp]
\begin{tabular}{cc}
\epsfig{file=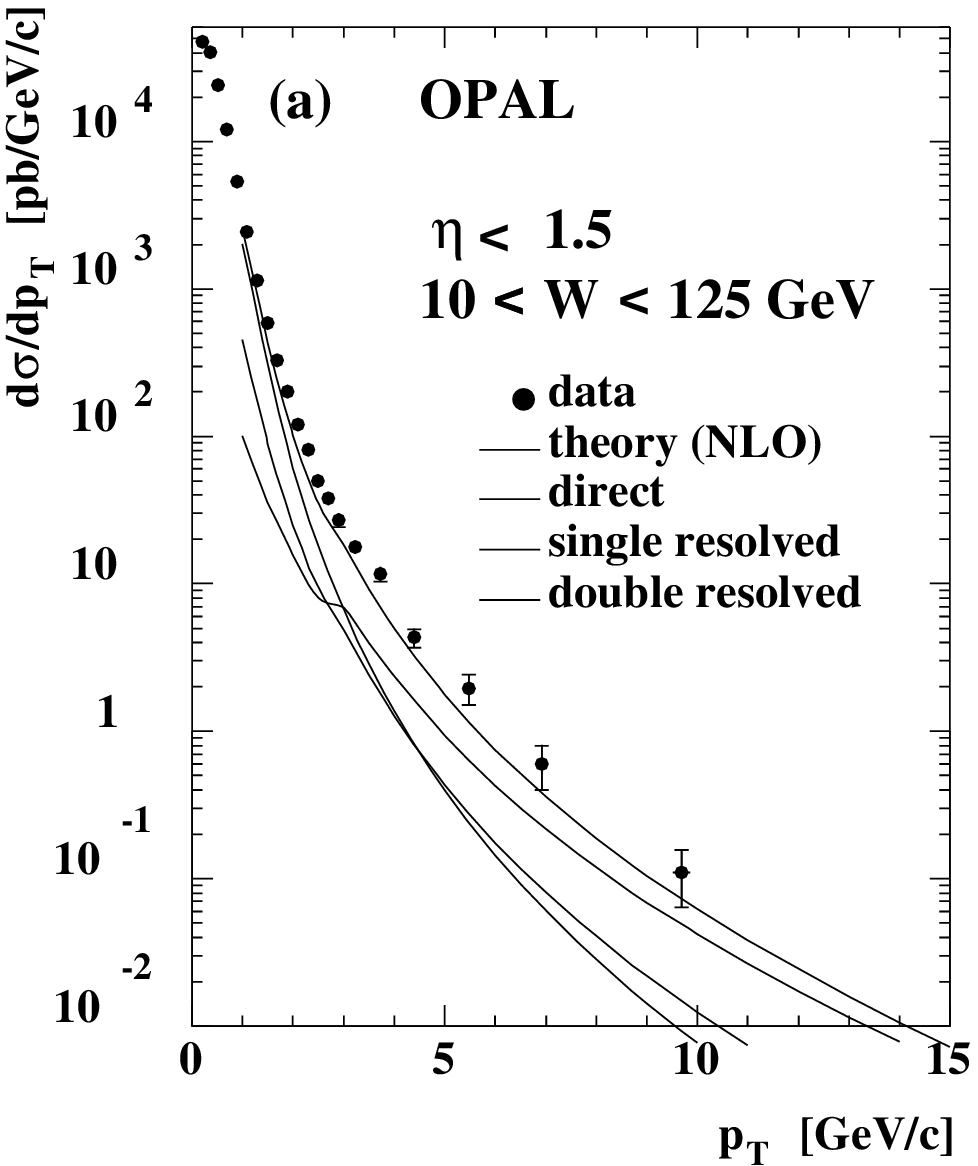,width=2.1in,height=1.87in} &
\epsfig{file=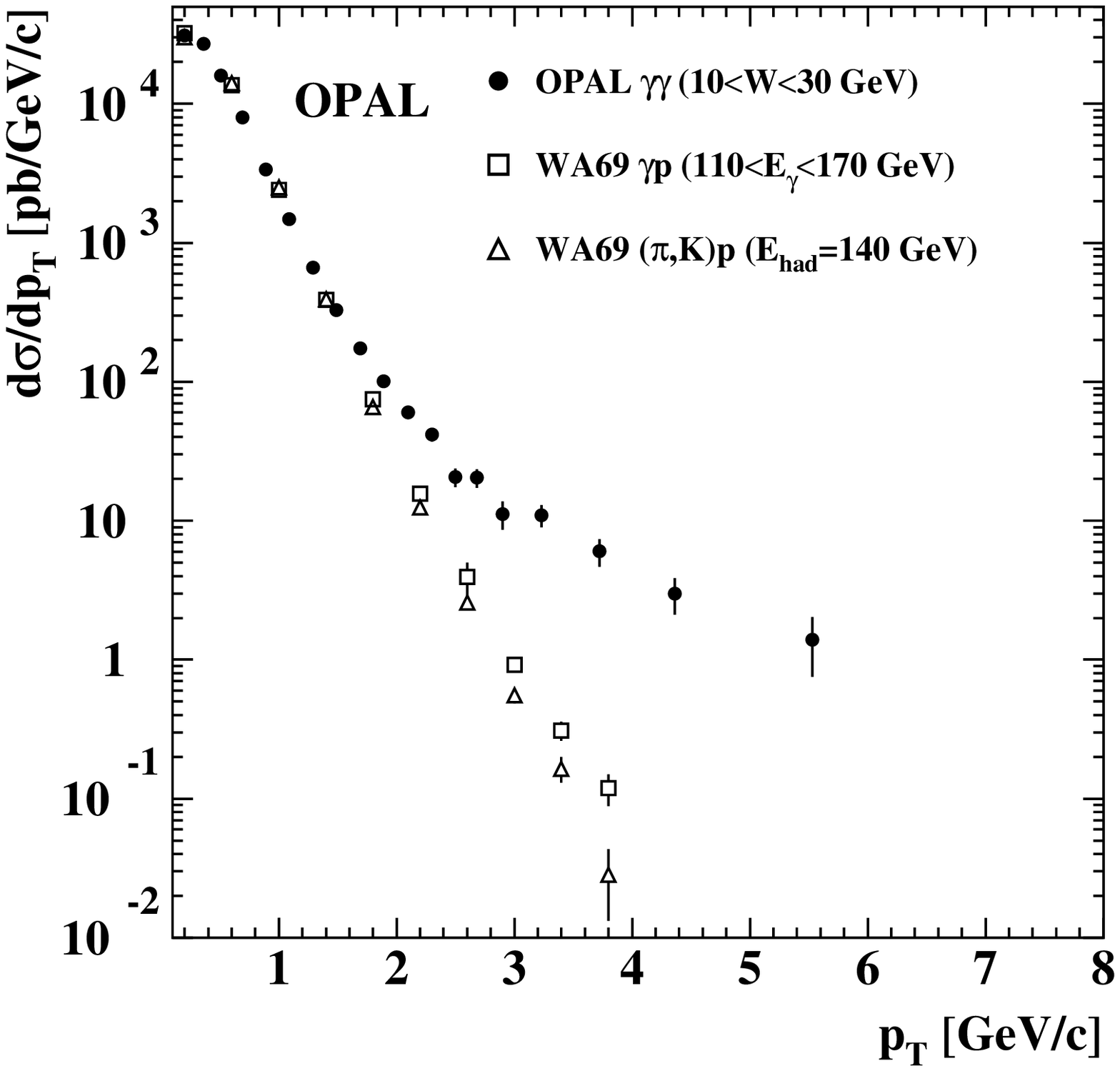,width=2.1in,height=1.87in}  \\
\end{tabular}
\caption{\it a) Differential inclusive charged hadron cross-section and b) the $\rm{p_{T}}$ distribution
measured in $\gg$ interactions compared to the $\gamma$p and ($\pi$,K)p interactions.}
\label{fig:single}
\end{figure}

Inclusive production of charged hadrons, $\rm{K^0_{s}}$ mesons, and jet studies has been performed 
at LEP by the OPAL experiment. Figure~\ref{fig:single}(a) shows a measurement of differential 
cross-section for charged hadrons produced in collision of the two quasi-real photons in the range 
10~GeV$< \wgg <$125~GeV as a function of transverse momentum\cite{chargopal} $p_{T}$ . The results are 
compared to NLO perturbative QCD calculations\cite{nlo}. For lower values of $\wgg$, more charged hadrons
\noindent  than predicted are 
found at large $p_{T}$. Also shown in figure~\ref{fig:single}(b) is the comparison of the $\gg$ data
\begin{wrapfigure}{r}{180pt}
\includegraphics*[bb=21 359 553 788,width=15pc]{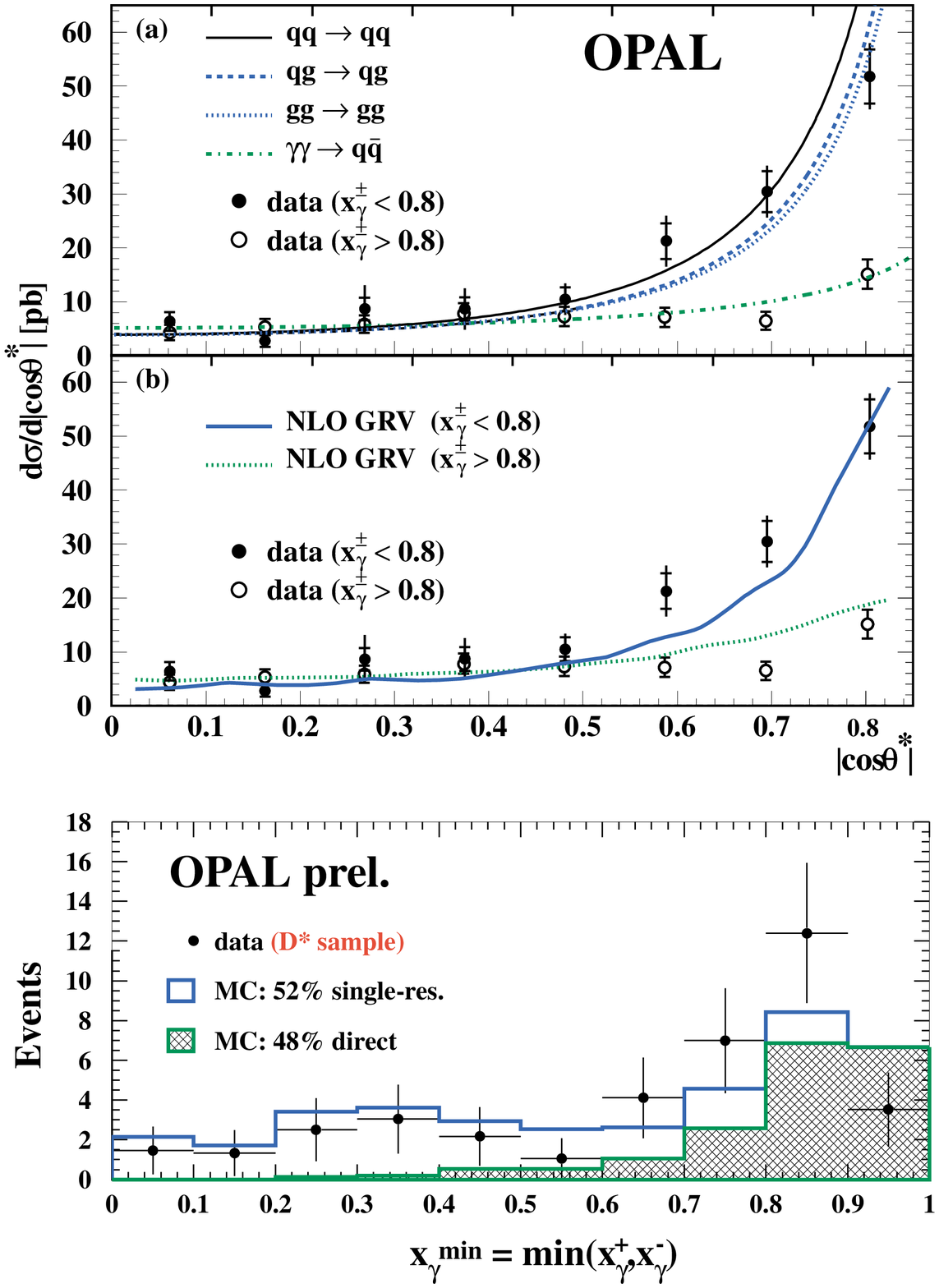}
\caption{\it The angular distribution in the dijet center-of-mass  system for "Direct" and 
"Resolved" events}
\label{fig:dijet}
\end{wrapfigure} 
data to $\mathrm{p_{T}}$ measured in $\mathrm{\gamma p}$ and 
($\mathrm{\pi,K)p}$ interactions normalised at the same value at low $p_{T}$, 
one observes there is a significant increase of rates in the $\gg$ process above a $p_{T}$ of 2~GeV. 
The clear deviation from the hadronic interactions shows the effect of the direct component in the 
$\gg$ interactions. Similar studies of $p_{T}$ distributions of the $K^0_{s}$ mesons are in reasonable 
agreement with NLO calculations\cite{chargopal}. 

The OPAL experiment has performed a very nice measurement of dijet production in two-photon
collisions at $\sqrt s = 161$ and 172 GeV. Their results\cite{opaldi} demonstrate  
that it is possible to distinguish between direct and resolved processes in the dijet events.
With the help of the variable $x_{\gamma}$, which is the estimator of the fraction of the target 
photon's momentum carried by the parton which produces jets. Figure~\ref{fig:dijet} shows 
the measured distribution of the parton scattering angle $\theta^*$ for direct and 
double-resolved processes, compared to the relevant QCD matrix element calculations\cite{dijet}. 
One observes a clear distinction between the direct process $\mathrm{\gg \rar \rm{q \bar{q}}} $ 
($x_{\gamma}>0.8 $), where a quark is exchanged in the t channel and the doubly resolved one ($ x_{\gamma}<0.8$), 
dominated by the gluon exchange. The strong rise in  $\mathrm{cos \theta^*}$ distribution near 
$\mathrm{cos \theta^*}$=1 is due to a large double-resolved contribution, as expected from QCD.

\section{Heavy Quark Production}
The study of heavy quark (c,b) production in two photon collisions at LEP provides not only 
an excellent test of perturbative QCD but also gives an estimate of the gluon density in the photon. 
At LEP energies, the direct and resolved photon processes are predicted to give comparable contributions 
to the charm and beauty quark production cross-sections\cite{drees}. The resolved process is dominantly
quark-gluon fusion: $\gamma$g $\rar q\bar{q}$. The cross-section of the processes 
$\mathrm{\ee \rar \ee c\bar{c}, b\bar{b}}$ X has been measured by the L3\cite{l3_charm} and 
OPAL\cite{opal_charm} experiments. At L3, the charm and beauty quark are identified by tagging 
leptons ($\mathrm{e , \mu}$) from semileptonic charm and beauty decays. Charm quark were also identified 
by the reconstruction of $\mathrm{D^{\pm *}}$ meson decays, where $\mathrm{D^{*} \rar D^0 \pi^{\pm}}$, and 
OPAL tags charm quark with $\mathrm{D^{*} \rar D^0 \pi^{\pm}}$ and $\mathrm {D^0 \rar K^{-} \pi^{+}, 
K^{-} \pi^{+} \pi^{o},K^{-} \pi^{+} \pi^{-}}$. 

\begin{figure}[htp]
  \begin{tabular}{lr}
  \includegraphics*[bb=21 77 533 336,height=9pc,width=13pc]{opal_jetcl.eps}&
  \includegraphics[height=9pc,width=13pc]{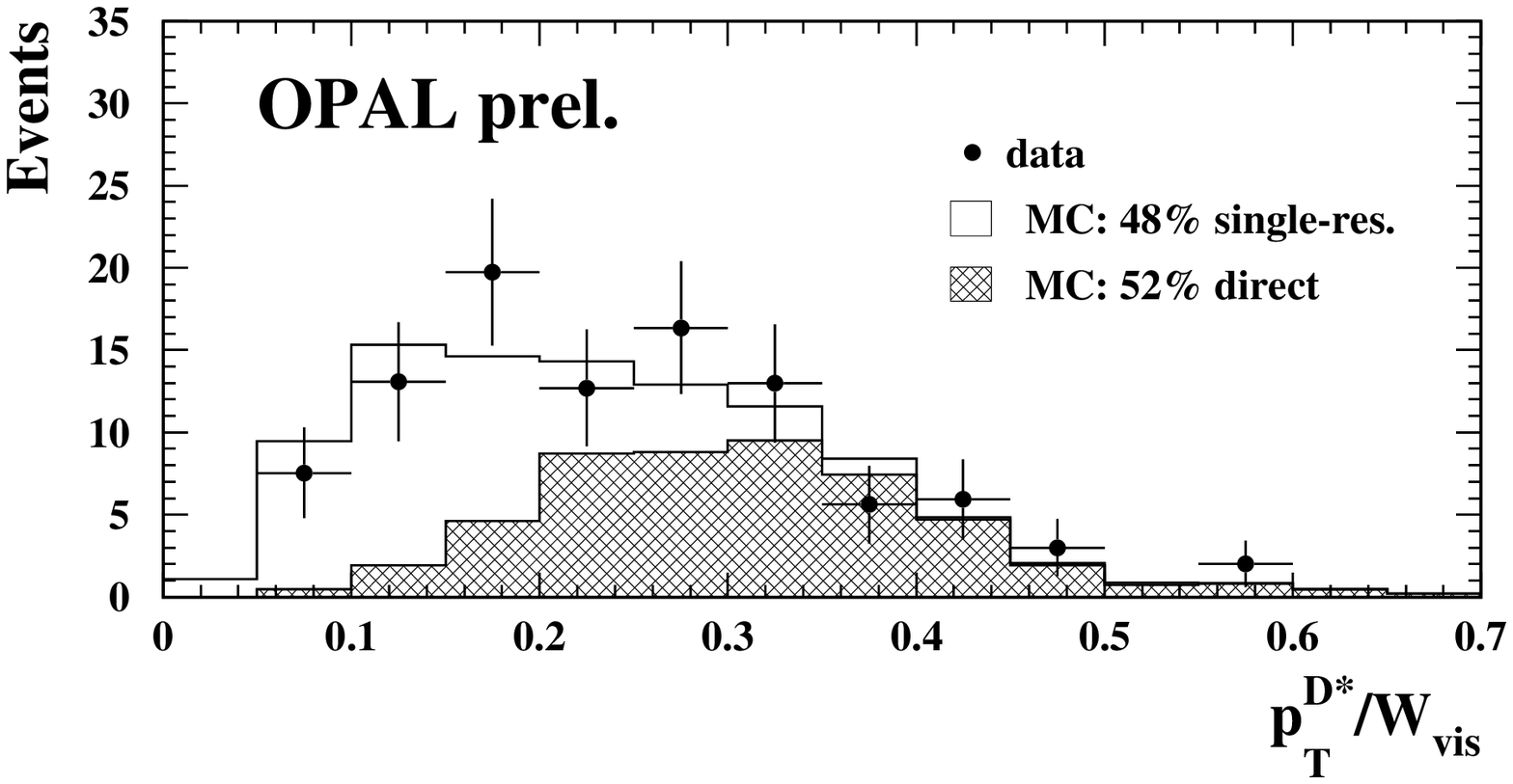}\\
\end{tabular}    
 \vspace*{-1pc}
  \caption{\it The $x_{\gamma}$ distribution of dijet events containing a $D^*$ 
   and the $p_{T}$  distribution of the $D^*$ normalised to the visible mass of 
   the event.}
  \label{fig:xgdstar}
\end{figure}
A good separation of direct and resolved processes is obtained by associating the $\mathrm{D^*}$
to a dijet analysis or by inspection of the $p_{T}$ distribution of the $\mathrm{D^*}$ 
(See figure~\ref{fig:xgdstar}). As predicted the direct and resolved processes contribute 
roughly equally to the observed distribution. The differential $\mathrm{D^*}$ cross section agrees
well with the NLO predictions and is independent of the Monte Carlo models used to correct the data 
over the range of detector acceptance. The 
\begin{wrapfigure}[18]{r}{180pt}
  \includegraphics[width=15pc]{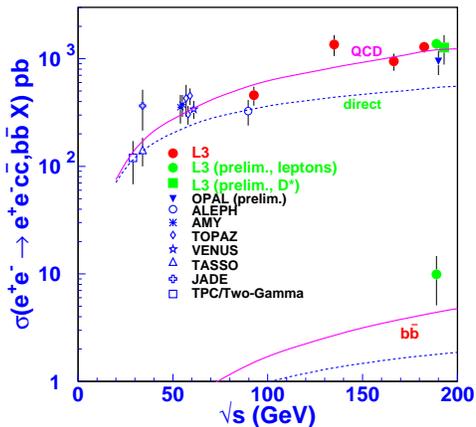}
  \caption[]{\it The  cross-section for heavy quarks
  production as measured at LEP and at previous $\ee$ colliders}
  \label{fig:charm}
\end{wrapfigure}
total inclusive cross-sections are plotted in Figure~\ref{fig:charm} together 
with previous measurements. The data are compared to NLO QCD calculations\cite{drees}. 
The direct process $\rm {\gamma \gamma \rar c\bar{c}, b\bar{b}}$, shown with dotted line, 
is insufficient to describe the data, even if real and virtual gluon corrections are included. 
The cross sections requires contributions from the resolved processes which are dominantly 
$ \rm{ \gamma g \rar c\bar{c}, b\bar{b}}$. The data therefore requires a significant gluon content 
in the photon.

The $\rm {b\bar{b}}$ cross section is measured for the first time in two photon collisions by the L3 
experiment. The preliminary value of b cross section lie somewhat above QCD predictions.

\section{Leptonic Structure Function, ${F_{2}^{\gamma, QED}}$}
The leptonic structure function has been measured by all LEP experiments\cite{af2qe,df2qe,l3f2qe,of2qe}.
The measurement provides not only a QED test but also an experimental check for the procedures 
used in the study of the hadronic photon structure functions. 

\begin{figure}[htp]
\begin{tabular}{ccc}
\epsfig{file=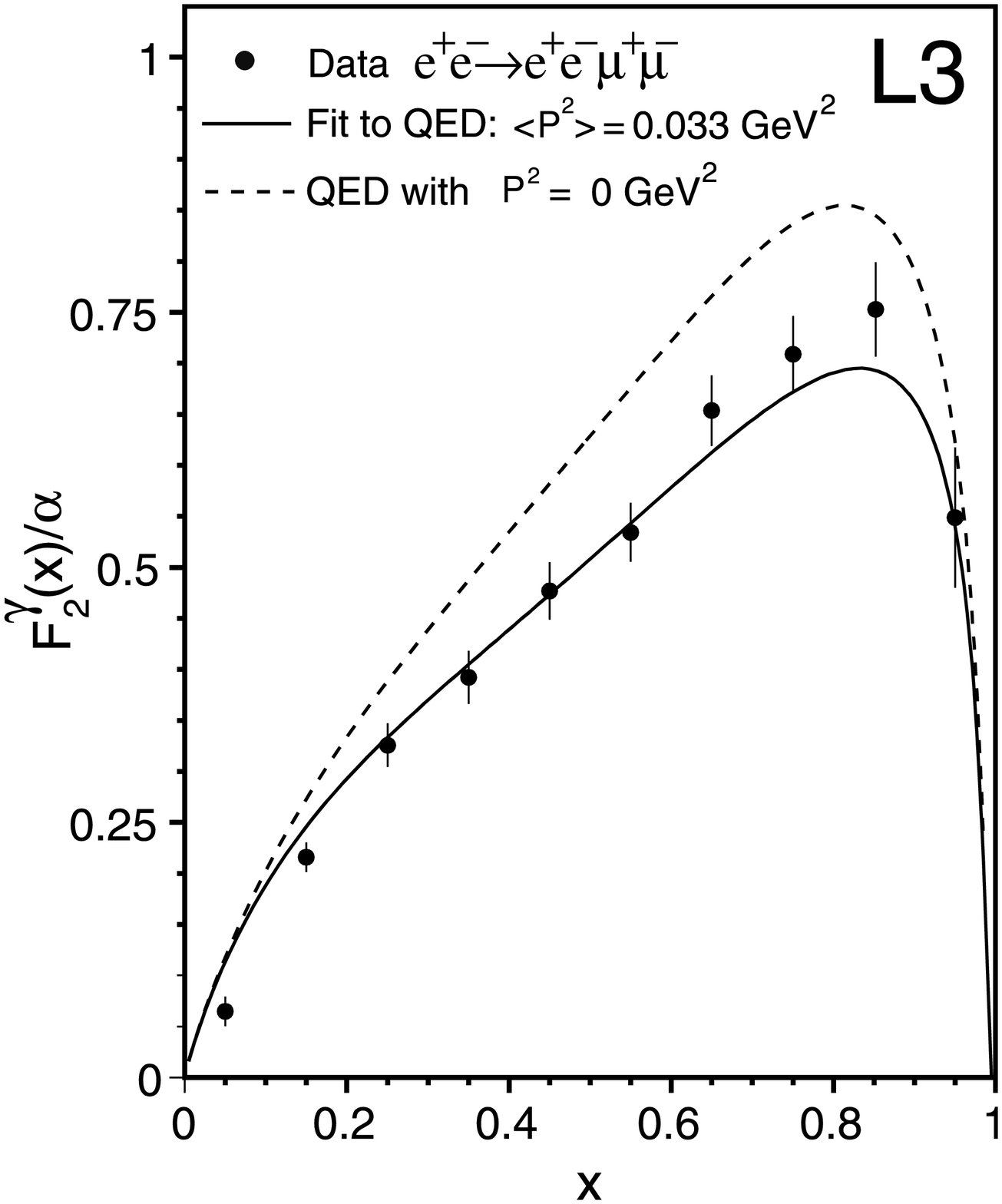,height=0.40\textwidth,width=0.30\textwidth} &
\epsfig{file=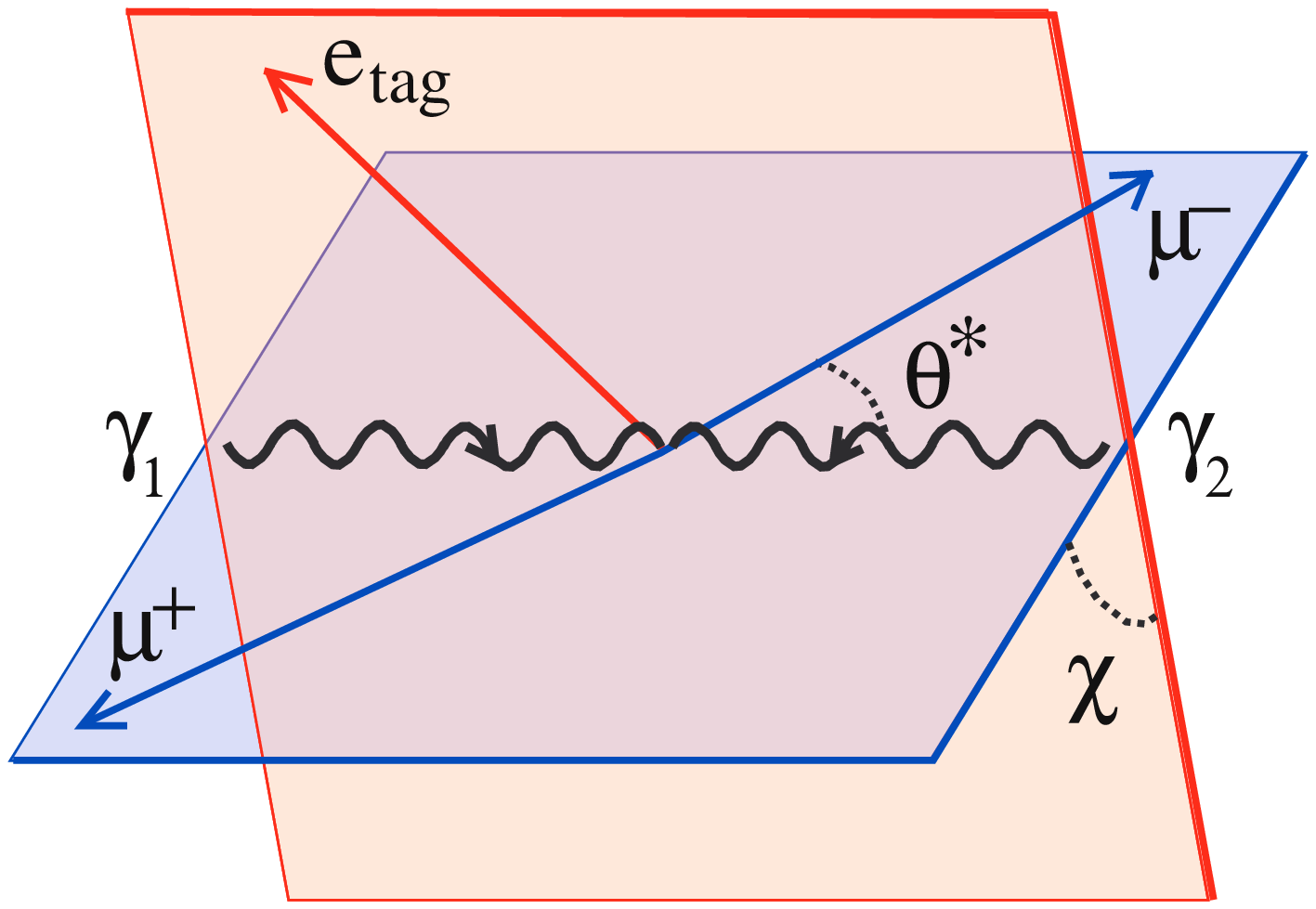,height=0.35\textwidth,width=0.30\textwidth}&
\epsfig{file=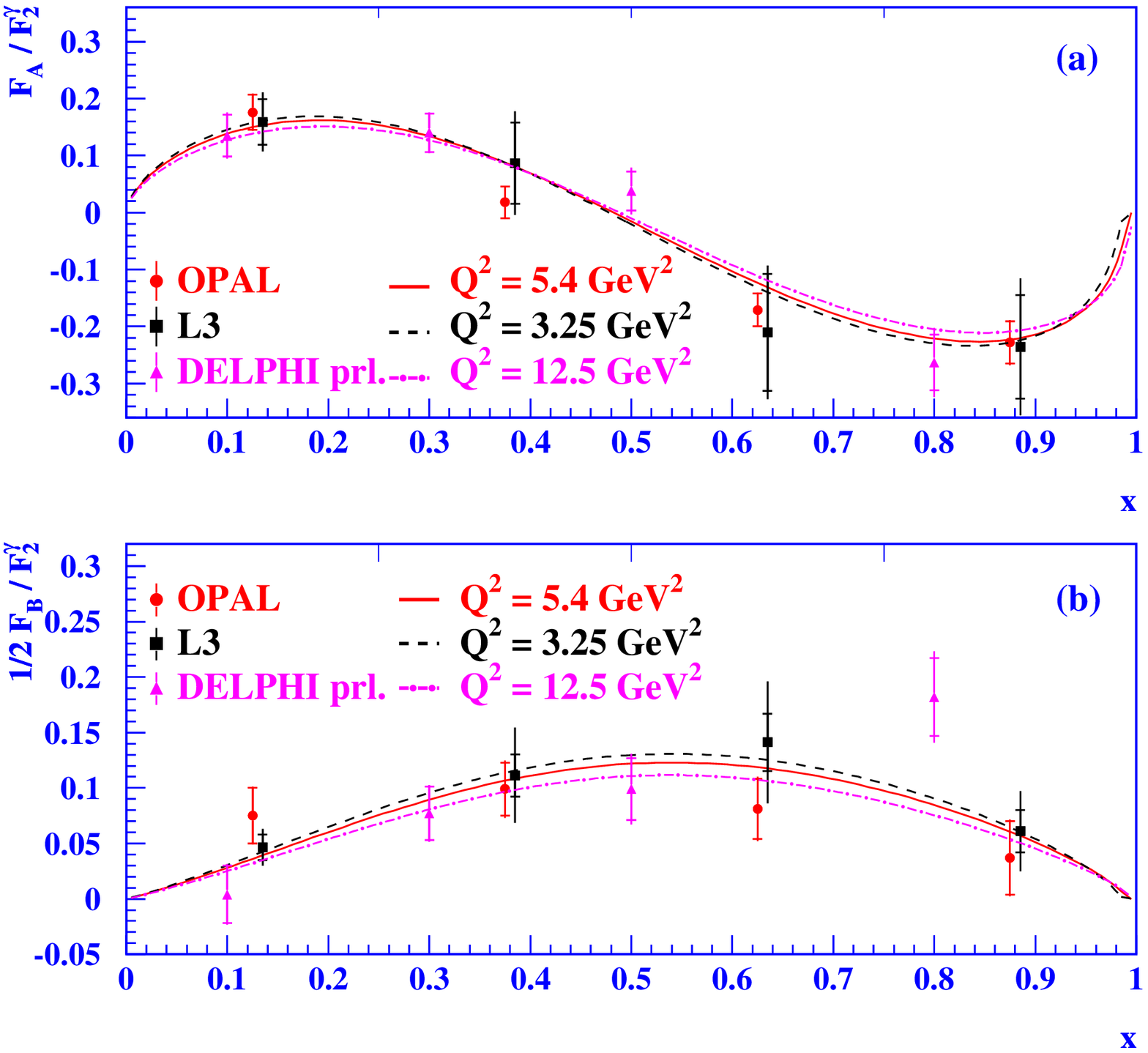,height=0.40\textwidth,width=0.30\textwidth} \\
(a) & (b) & (c)
\end{tabular}
\caption{\it (a) $\fg$ measured in the range $1.4 < \rm{Q^2} < 7.6~\gev$, (b) definition of the
angles $\theta^*$ and $\chi$ in the $\gg$ centre of mass frame and (c) measurement of the $F_A$
and $F_B$ structure of function. }
\label{fig:f2qed}
\end{figure}

A result from L3, is shown as an example in figure~\ref{fig:f2qed} (a). It shows that it is 
possible to measure the effect of non-zero target photon virtuality. The analysis is performed 
using  the $\ee \rar \mm$ sample, for a range of $\mathrm{Q^2}$ ($1.4 < \mathrm{Q^2} < 7.6~\rm{GeV}^{2}$).  
The fit to $F_{2}^{\gamma, QED}$ corresponds to a target photon virtuality of $\rm{0.33 \pm 0.005~GeV^{2}}$, 
in good agreement with QED predictions, if initial state radiative corrections are included.

Also shown in Figure~\ref{fig:f2qed} (c), is the measurement of the $F_A$ and $F_B$
structure functions , obtained by studying the azimuthal angle distribution of the $\mu ^-$ in 
the $\gamma \gamma$ centre-of-mass system\cite{cello,azimuth,yellow,peterson,nisius}. Assuming that the target 
photon direction is parallel to the beam axis, the polar angle $\theta^*$ of the  
$\mu^-$ and  the azimuthal angle $\chi$ are defined as shown in
Figure~\ref{fig:f2qed}(b). Here $\chi$ is the angle between the plane defined by the $\mu^-$ direction 
and the $\gamma \gamma$ axis, and the scattering  plane of the tagged electron. Both structure 
functions $F_A$ and $F_B$, originate from the interference terms of the scattering amplitudes. The  
characteristic $x$ dependence of the interference terms, as predicted by QED, is observed in the 
data as shown in figure~\ref{fig:f2qed} (c). In particular $F_A$ is due to the interference between 
longitudinal-transverse and transverse-transverse photon amplitudes, thus providing information on the 
longitudinal component of the probe photon. With this measurement, LEP proves  that the longitudinal 
leptonic photon helicity amplitude can be accessed by the study of azimuthal correlations and is 
significantly non-zero. 

\section{Hadronic structure function $\f2qcd$}
\begin{figure}[htp]
\begin{tabular}{cc}
\epsfig{file=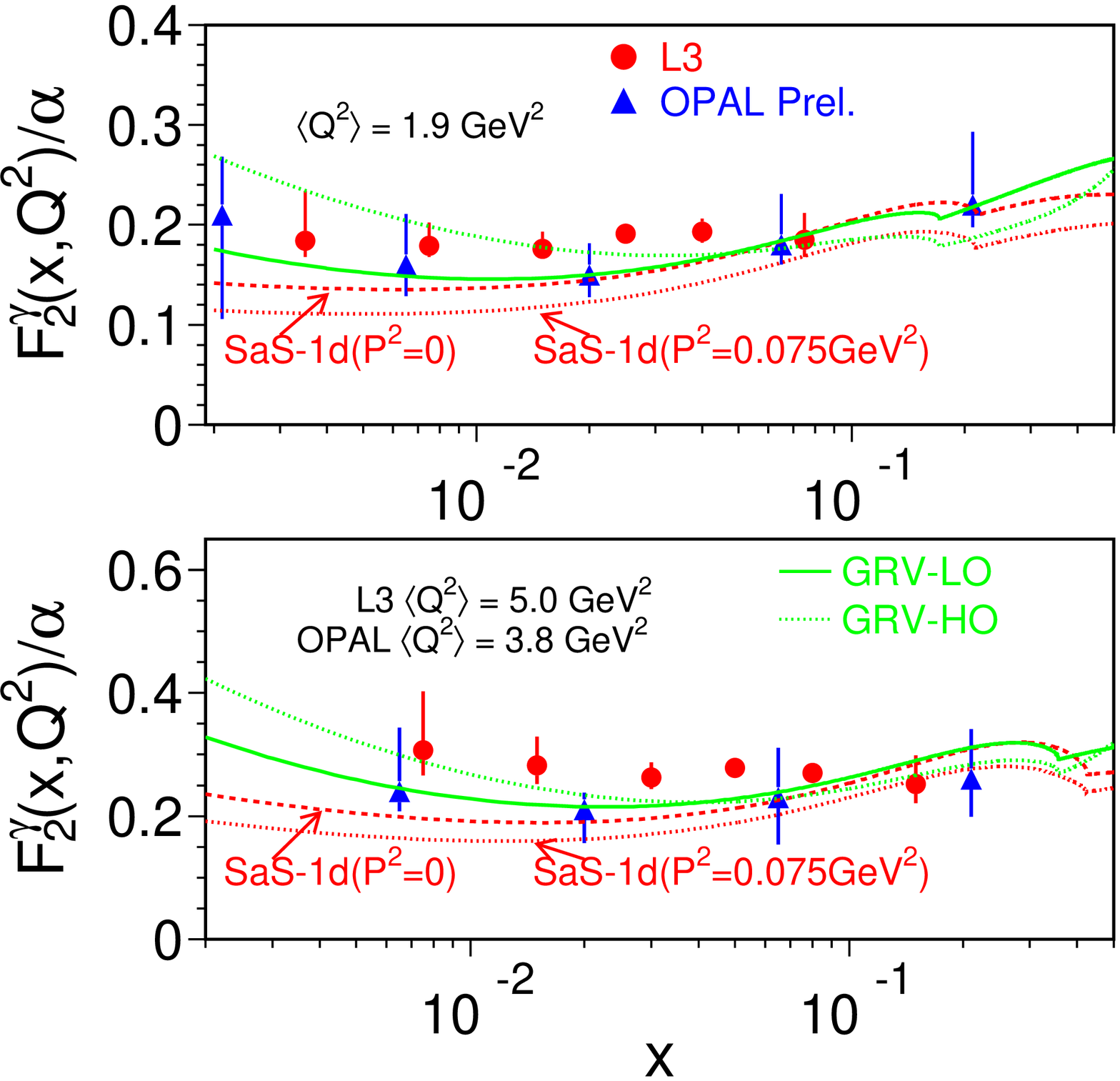,width=0.45\textwidth,height=0.45\textheight}&
\epsfig{file=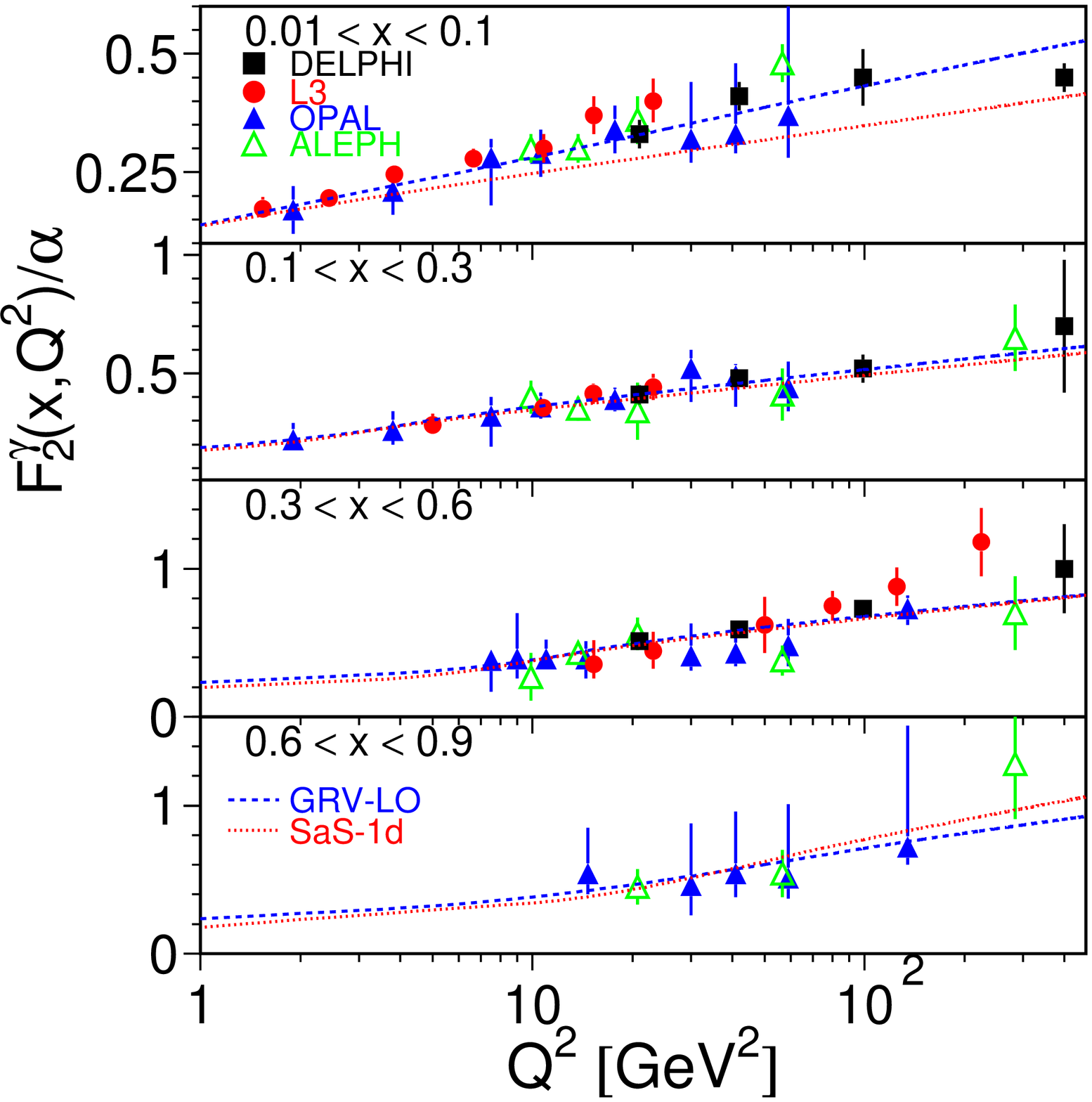,width=0.45\textwidth,height=0.45\textheight} \\
(a) &  (b)
\end{tabular}
\caption{\it (a) The measured $\fg$ at $\langle Q^2 \rangle = 1.9$~GeV$^2$ and 5.0~GeV$^2$ and
         (b) evolution of $\fg$ as a function of $Q^2$ for different range of $x$ values.  }

\label{fig:f2qall}
\end{figure}

The measurement of the hadronic structure function, $\f2qcd$ has been performed at LEP in the range 
$ 0.0025< x < 1$  and $ 1.2~\gev < Q^{2} < 279~\gev$ \cite{af2q,df2q,l3f2q,of2q}. 
The physical interest in the analysis of the hadronic photon structure function is twofold. 
Firstly, to measure the shape of $\fg$, especially at small values of $x$, at fixed $Q^{2}$, 
where HERA experiments observe a strong rise of the proton structure function. Secondly the 
$\rm{Q^{2}}$ evolution of $\fg$ is investigated. The $\fg$ measurements from L3 and OPAL are 
shown in Figure~\ref{fig:f2qall} (a) in the $Q^2$ interval from 1.2 to 9.0~GeV$^2$. The $x$ range 
is $0.002< x < 0.1$ at $\langle Q^2 \rangle = 1.9$~GeV$^2$ and $0.005 < x < 0.2$ at
$\langle Q^2 \rangle = 5.0$~GeV$^2$. For the low values of $x$, the data agree better with the 
parton density prediction of GRV\cite{grv}, whereas SaS-1d\cite{sas} prediction is lower.

A compilation of the results for different experiments on the $Q^2$ evolution of $\fg$ in various 
ranges of $x$ are shown in  Figure~\ref{fig:f2qall} (b). The measured values of $\fg$ show clearly 
the linear growth with $\ln Q^2$ expected by QCD. The predictions of the  GRV-LO\cite{grv} and 
SaS-1d\cite{sas} models are also shown. With all the statistics available at the end of LEP data 
taking, one hopes to extract the effective scale parameter $\Lambda_{QCD}$ at large x.

\section{$\gamma^{*} \gamma^{*}$ Collisions}
\begin{figure}[htp]
\begin{tabular}{cc}
\epsfig{file=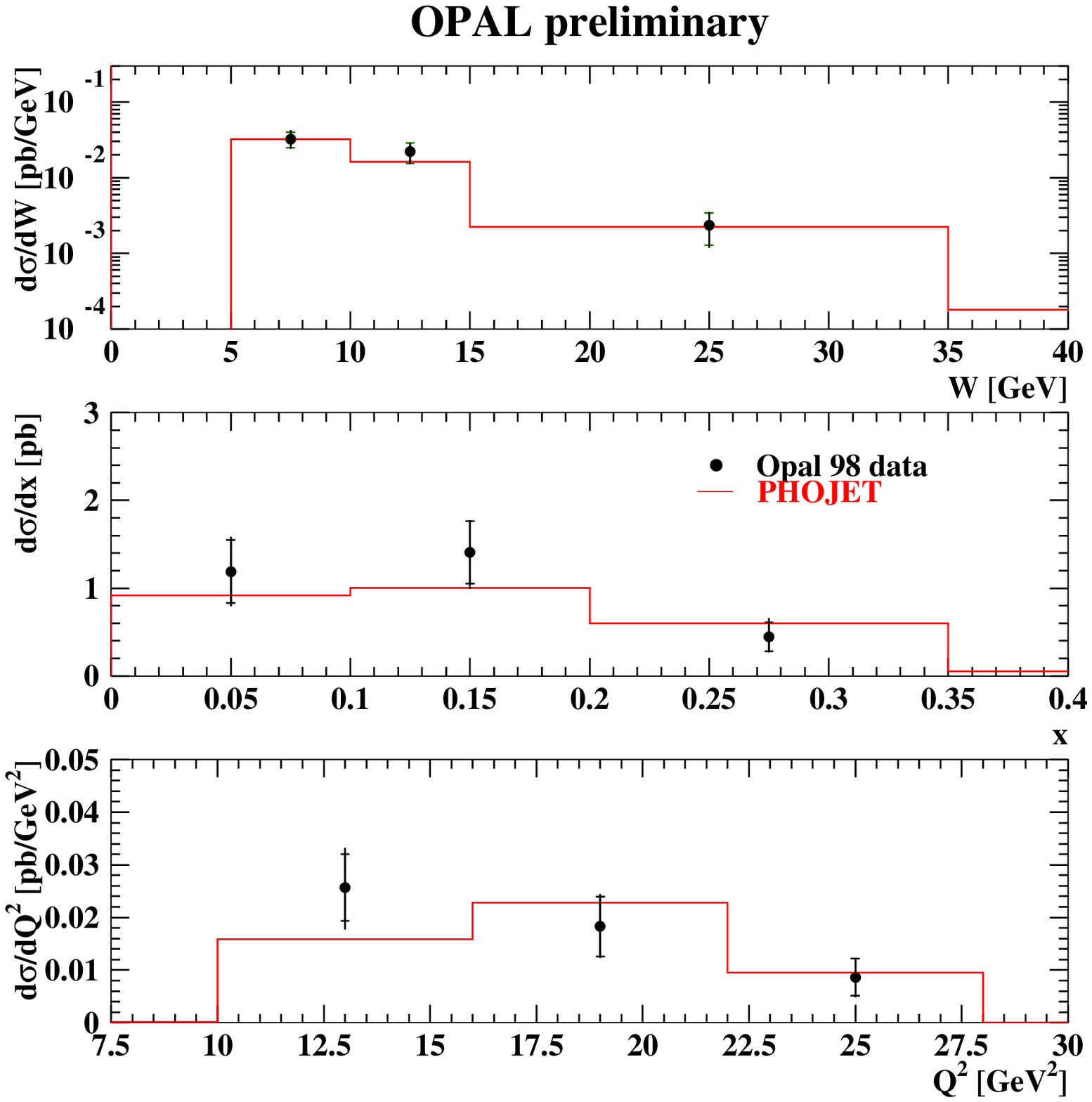,width=0.45\textwidth,height=0.43\textheight}&
\epsfig{file=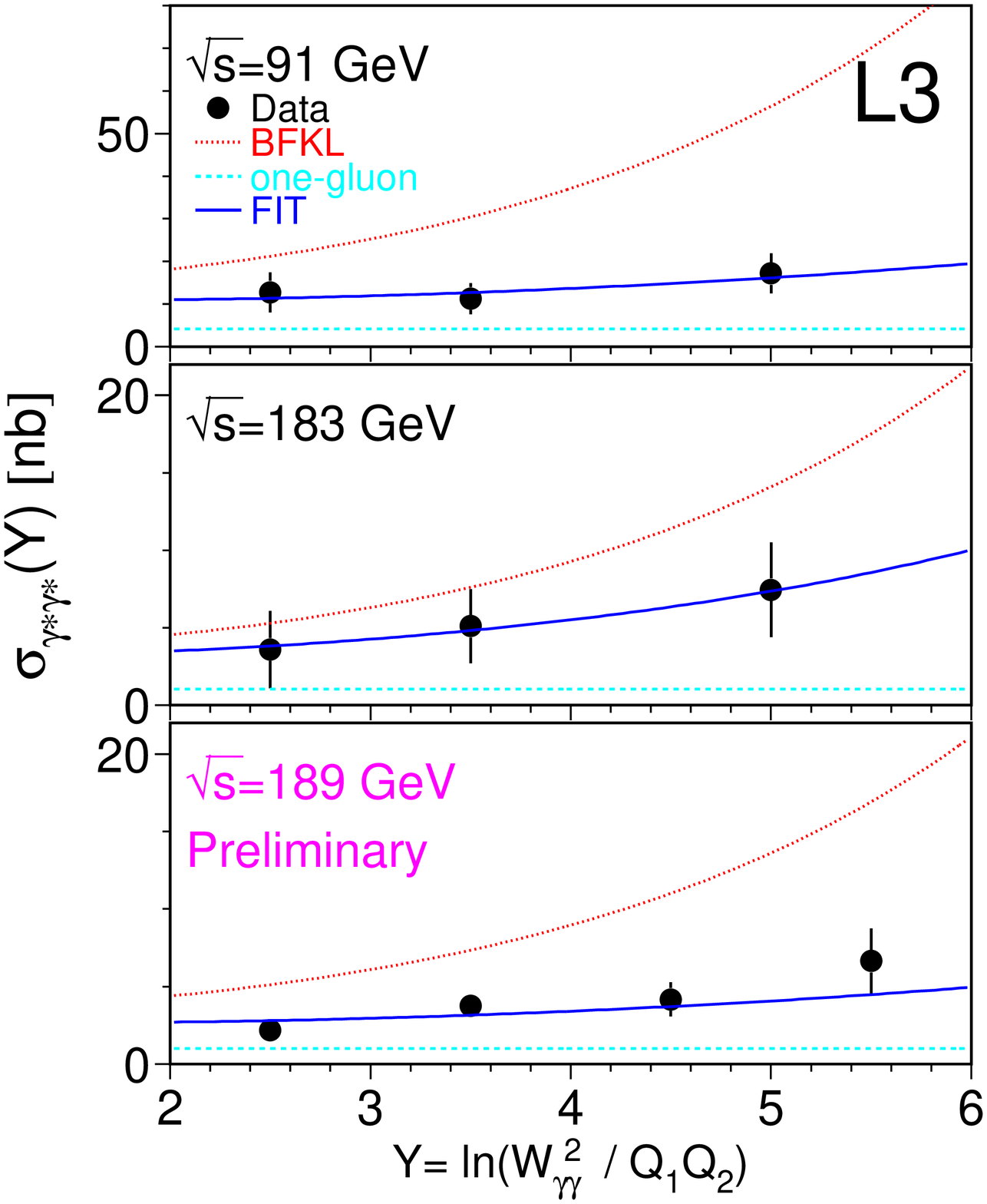,width=0.45\textwidth,height=0.43\textheight} \\
(a) &  (b)
\end{tabular}
\caption{\it (a) The differential cross-section of double tag events compared to PHOJET Monte Carlo predictions 
             and (b) the two photon cross-sections at LEP1 and LEP2 compared to LO-BFKL calculations after
	     subtraction of the direct contribution }

\label{fig:bfkl}
\end{figure}

The cross-section of $\gamma ^* \gamma ^*$ collisions has been measured 
at LEP with L3\cite{l3_bfkl} and OPAL\cite{opal_bfkl} experiments in the range of 
$3~\rm{GeV^2} < \rm{Q^2_{1,2}} < 37~\rm{GeV^2}$. Since the two photons are highly virtual and unlike the proton, 
they do not contain constituent quarks with an unknown density distribution, so one may hope to have 
a complete perturbative QCD calculation under particular kinematical conditions. An alternative QCD approach 
is based on the BFKL equation\cite{bfkl}. Here the highly virtual two-photon process, 
with $Q^2_1\simeq Q^2_2$, is considered as the ``golden'' process where the calculation can be 
verified without phenomenological inputs\cite{gg2,gg1}. The $\gamma ^* \gamma^*$ interaction can 
be seen as the  interaction of two $\rm{q \bar{q}}$ pairs scattering off each other via multiple gluon 
exchange. In this scheme the cross-section for the collision of two virtual photons \cite{gg2,gg1}
depends upon the ``hard Pomeron'' intercept $\alpha_P -1 = 0.53$\cite{gg2,gg1} in the LO, whereas
in the next-to-leading order the BFKL contribution is calculated to be smaller,
$\alpha_P -1 \simeq 0.17$\cite{gg3}. The results from L3 and OPAL (figure~\ref{fig:bfkl}(a)) show that 
the events are well described by the PHOJET Monte Carlo model which uses the GRV-LO parton density in the 
photon and leading order perturbative QCD. The LO BFKL calculations shown in the figure~\ref{fig:bfkl}(b)  
with dotted line are too high. By leaving $\alpha_P$ as a free parameter in the LO calculations, a combined fit to 
the L3 

\noindent data obtained at $\sqrt{s} \approx 91,$ 183 and 189~GeV gives a value of $\alpha_P - 1 = 0.29 \pm 0.025$ 
with $\chi^2$/d.o.f =7/9.

\section*{Outlook}
Progress in the field of the two photon physics at LEP is significant, most notable are 
multi-hadron production and photon structure functions. With the statistics of $500~\mathrm{pb}^{-1}$ 
luminosity available at the end of LEP II data taking, we expect not only large improvements in the 
understanding of the photon structure function at small $x$ values but also have possibility to actually
observe glueball states with very low two photon widths. 

\section*{Acknowledgements}

I would like to thank all the LEP colleagues for their contribution. I am grateful to J.H. Field and 
M.N. Kienzle-Focacci for encouraging discussions. This work is supported by the Swiss National Science Foundation.

\end{document}